# A Block Minorization–Maximization Algorithm for Heteroscedastic Regression

Hien D. Nguyen[12], Luke R. Lloyd-Jones[3], and Geoffrey J. McLachlan[1*]

May 30, 2016

**Abstract**

The computation of the maximum likelihood (ML) estimator for heteroscedastic regression models is considered. The traditional Newton algorithms for the problem require matrix multiplications and inversions, which are bottlenecks in modern Big Data contexts. A new Big Data-appropriate minorization–maximization (MM) algorithm is considered for the computation of the ML estimator. The MM algorithm is proved to generate monotonically increasing sequences of likelihood values and to be convergent to a stationary point of the log-likelihood function. A distributed and parallel implementation of the MM algorithm is presented and the MM algorithm is shown to have differing time complexity to the Newton algorithm. Simulation studies demonstrate that the MM algorithm improves upon the computation time of the Newton algorithm in some practical scenarios where the number of observations is large.

---

[*1]School of Mathematics and Physics, University of Queensland (Email: h.nguyen7@uq.edu.au; g.mclachlan@uq.edu.au). [2]Centre for Advanced Imaging, University of Queensland. [3]Queensland Brain Institute, University of Queensland (Email: l.lloydjones@uq.edu.au).

# 1 Introduction

One of the major challenges of the analysis of Big Data is the requirement to make fundamental and standard statistical processes applicable in the presence of the various computational challenges; see [1, 2] for details. One major theme of Big Data research is to construct algorithms for standard statistical processes that allow for parallelization and distributed computing.

Let $Y_1, ..., Y_n$ be an independent and identically distributed sample such that $Y_i$ is normal with mean $\mu_i(\boldsymbol{x}_i)$ and variance $\sigma_i^2(\boldsymbol{x}_i)$ for $i = 1, ..., n$, where $\boldsymbol{x}_i^T = (x_{i1}, ..., x_{id}) \in \mathbb{R}^d$ is a vector of covariates. The superscript $T$ indicates matrix transposition. Suppose that we observe the realizations $y_1, ..., y_n$ and wish to estimate the parametric mean and variance functions $\mu_i(\boldsymbol{x}_i) = \boldsymbol{\beta}^T \boldsymbol{x}_i$ and $\sigma_i^2(\boldsymbol{x}_i) = \sigma^2$, respectively, where $\boldsymbol{\beta}^T = (\beta_1, ..., \beta_d) \in \mathbb{R}^d$ and $\sigma^2 > 0$. The characterization describes the standard linear regression model, and the parameter vector of the model, $\boldsymbol{\theta}^T = (\boldsymbol{\beta}, \sigma^2)$, can be estimated via the maximum likelihood (ML) estimator

$$\hat{\boldsymbol{\theta}} = \arg\max_{(\boldsymbol{\beta}, \sigma^2)} \prod_{i=1}^{n} \phi\left(y_i; \boldsymbol{\beta}^T \boldsymbol{x}_i, \sigma^2\right), \tag{1}$$

where

$$\phi\left(y, \mu, \sigma^2\right) = \left(2\pi\sigma^2\right)^{-1/2} \exp\left(-[y-\mu]^2 / [2\sigma^2]\right)$$

is the normal density function in $y$, with mean $\mu$ and variance $\sigma^2$. Based on the minorization–maximization framework (MM; see [3] for details), [4] constructed a simple algorithm for computing (1) that allows for parallelization and distributed computing.

In this letter, we consider the case where the parametric mean and variance functions are characterized by $\mu_i(\boldsymbol{x}_i) = \boldsymbol{\beta}^T \boldsymbol{x}_i$ and $\sigma_i^2(\boldsymbol{x}_i) = \exp(\boldsymbol{\alpha}^T \boldsymbol{x}_i)$, where $\boldsymbol{\alpha}^T = (\alpha_1, ..., \alpha_d) \in \mathbb{R}^d$. The characterization describes the multiplicative het-eroscedasticity regression model of [5]; see also [6] and [7, Sec. 11.7]. Recent applications of such model include modeling of loans in empirical finance [8], econometric time-series analysis [9], and whole-genome prediction [10]. The parameter vector of the model, $\boldsymbol{\psi}^T = \left(\boldsymbol{\alpha}^T, \boldsymbol{\beta}^T\right)$, can be estimated via the ML estimator

$$\hat{\boldsymbol{\psi}} = \arg\max_{(\boldsymbol{\alpha}, \boldsymbol{\beta})} \prod_{i=1}^{n} \phi\left(y_i; \boldsymbol{\beta}^T \boldsymbol{x}_i, \exp\left[\boldsymbol{\alpha}^T \boldsymbol{x}_i\right]\right). \qquad (2)$$

The ML estimator (2) can be computed via a Newton algorithm [7, Sec. 11.7], which is not appropriate for parallelization and distributed computing. This is due to the need for the repeated inversion of $d \times d$ matrices, which can both be large or numerically singular. In a recent review of Big Data algorithms, [11] presented no specialized algorithms or software for regression under heteroscedasticity. We extend upon the work of [4] to produce a Big Data-appropriate MM algorithm for the computation of (2), using the recent developments in geometric and signomial programming of [12]. Although it is possible to parallelize the Newton algorithm via matrix parallelization techniques such as those that are discussed in [13, 14, 15], our algorithm permits a more intuitive and simpler implementation.

We use recent results from optimization in signals processing [16, 17] to establish global convergence results for the derived algorithm. We also briefly study the time complexity of the MM algorithm, and outline a framework for a parallel and distributed implementation of the algorithm. Simulation studies are conducted to demonstrate the computational performance of the algorithm in both serial and parallel implementations. Comparisons between the MM algorithm and the Newton algorithm are made.

## 2 The MM Algorithm

The ML estimator (2) can be rewritten as

$$\hat{\boldsymbol{\psi}} = \arg\max_{(\boldsymbol{\alpha},\boldsymbol{\beta})} \ell(\boldsymbol{\alpha},\boldsymbol{\beta}), \qquad (3)$$

where

$$\begin{aligned} \ell(\boldsymbol{\alpha},\boldsymbol{\beta}) &= \sum_{i=1}^{n} \log \phi\left(y_i; \boldsymbol{\beta}^T \boldsymbol{x}_i, \exp\left[\boldsymbol{\alpha}^T \boldsymbol{x}_i\right]\right) \qquad (4) \\ &= -\frac{n \log(2\pi)}{2} - \sum_{i=1}^{n} \frac{\boldsymbol{\alpha}^T \boldsymbol{x}_i}{2} - \sum_{i=1}^{n} \frac{(y_i - \boldsymbol{x}_i^T \boldsymbol{\beta})^2}{2 \exp(\boldsymbol{x}_i^T \boldsymbol{\alpha})} \end{aligned}$$

is the log-likelihood function. Define the blockwise minorizer of $\ell(\boldsymbol{\alpha},\boldsymbol{\beta})$ at the point $\tilde{\boldsymbol{\psi}}^T = \left(\tilde{\boldsymbol{\alpha}}^T, \tilde{\boldsymbol{\beta}}^T\right)$, in the $\boldsymbol{\alpha}$ block, as a function $Q_{\boldsymbol{\alpha}}\left(\boldsymbol{\alpha}; \tilde{\boldsymbol{\psi}}\right)$ for $\boldsymbol{\alpha} \in \mathbb{R}^d$, with the properties that (i) $\ell\left(\tilde{\boldsymbol{\alpha}}, \tilde{\boldsymbol{\beta}}\right) = Q_{\boldsymbol{\alpha}}\left(\tilde{\boldsymbol{\alpha}}; \tilde{\boldsymbol{\psi}}\right)$ and (ii) $\ell\left(\boldsymbol{\alpha}, \tilde{\boldsymbol{\beta}}\right) \geq Q_{\boldsymbol{\alpha}}\left(\boldsymbol{\alpha}; \tilde{\boldsymbol{\psi}}\right)$; the blockwise minorizer in the $\boldsymbol{\beta}$ block is defined as $Q_{\boldsymbol{\beta}}\left(\boldsymbol{\beta}; \tilde{\boldsymbol{\psi}}\right)$ for $\boldsymbol{\beta} \in \mathbb{R}^d$, with (i) and (ii) replaced by $\ell\left(\tilde{\boldsymbol{\alpha}}, \tilde{\boldsymbol{\beta}}\right) = Q_{\boldsymbol{\beta}}\left(\tilde{\boldsymbol{\beta}}; \tilde{\boldsymbol{\psi}}\right)$ and $\ell(\tilde{\boldsymbol{\alpha}}, \boldsymbol{\beta}) \geq Q_{\boldsymbol{\beta}}\left(\boldsymbol{\beta}; \tilde{\boldsymbol{\psi}}\right)$ for $\boldsymbol{\beta} \in \mathbb{R}^d$, respectively.

Let $\boldsymbol{\psi}^{(0)}$ be an initial value; a blockwise MM algorithm for computing (3) is defined via the update scheme

$$\boldsymbol{\psi}^{(r+1)T} = \begin{cases} \left(\arg\max_{\boldsymbol{\alpha}} Q_{\boldsymbol{\alpha}}\left(\boldsymbol{\alpha}; \boldsymbol{\psi}^{(r)}\right), \boldsymbol{\beta}^{(r)}\right) & \text{if } r \text{ is odd}, \\ \left(\boldsymbol{\alpha}^{(r)}, \arg\max_{\boldsymbol{\beta}} Q_{\boldsymbol{\beta}}\left(\boldsymbol{\beta}; \boldsymbol{\psi}^{(r)}\right)\right) & \text{if } r \text{ is even}, \end{cases}$$

where $\boldsymbol{\psi}^{(r)T} = \left(\boldsymbol{\alpha}^{(r)T}, \boldsymbol{\beta}^{(r)T}\right)$ is the $r$th iterate of the algorithm. The following propositions provides $\boldsymbol{\alpha}$- and $\boldsymbol{\beta}$-blockwise minorizers for $\ell$.

**Proposition 1.** *Given $\boldsymbol{\psi}^{(r)}$, the log-likelihood (4) can be $\boldsymbol{\alpha}$-blockwise minorized*

by the minorizer

$$
\begin{aligned}
&Q_{\boldsymbol{\alpha}}\left(\boldsymbol{\alpha};\boldsymbol{\psi}^{(r)}\right)\\
=\ &-\frac{n\log(2\pi)}{2}-\frac{1}{2}\sum_{i=1}^{n}\sum_{j=1}^{d}\alpha_j x_{ij}\\
&-\frac{1}{2d}\sum_{i=1}^{n}\sum_{j=1}^{d}\frac{\left(y_i-\boldsymbol{x}_i^T\boldsymbol{\beta}^{(r)}\right)^2}{\exp\left(dx_{ij}\left(\alpha_j-\alpha_j^{(r)}\right)+\boldsymbol{x}_i^T\boldsymbol{\alpha}^{(r)}\right)}.
\end{aligned} \qquad (5)
$$

**Proposition 2.** *Given $\boldsymbol{\psi}^{(r)}$, the log-likelihood (4) can be $\boldsymbol{\beta}$-blockwise minorized by the minorizer*

$$
\begin{aligned}
&Q_{\boldsymbol{\beta}}\left(\boldsymbol{\beta};\boldsymbol{\psi}^{(r)}\right)\\
=\ &-\frac{n\log(2\pi)}{2}-\frac{1}{2}\sum_{i=1}^{n}\sum_{j=1}^{d}\alpha_j^{(r)} x_{ij}\\
&-\frac{1}{2d}\sum_{i=1}^{n}\sum_{j=1}^{d}\frac{\left(y_i-dx_{ij}\left(\beta_i-\beta_i^{(r)}\right)-\boldsymbol{x}_i^T\boldsymbol{\beta}^{(r)}\right)^2}{\exp\left(\boldsymbol{x}_i^T\boldsymbol{\alpha}^{(r)}\right)}.
\end{aligned} \qquad (6)
$$

Propositions 1 and 2 are adapted from [12, Eqn. 7] and [4, Eqn. 4.5]. We observe that (6) is linearly separable in $\beta_j$ for $j=1,...,d$, and that each $\beta_j$ occurs within a concave quadratic expression. Thus, (6) is concave in $\boldsymbol{\beta}$ and we solve the first-order condition equation $\nabla Q_{\boldsymbol{\beta}}\left(\boldsymbol{\beta};\boldsymbol{\psi}^{(r)}\right)=\mathbf{0}$ to obtain

$$\boldsymbol{\beta}^*=\arg\max_{\boldsymbol{\beta}} Q_{\boldsymbol{\beta}}\left(\boldsymbol{\beta};\boldsymbol{\psi}^{(r)}\right),$$

where $\nabla$ is the gradient operator, $\mathbf{0}$ is a vector of zeros, $\boldsymbol{\beta}^{*T}=(\beta_1^*,...,\beta_d^*)$, and

$$\beta_j^*=\beta_j^{(r)}+\frac{\sum_{i=1}^{n}x_{ij}\left(y_i-\boldsymbol{x}_i^T\boldsymbol{\beta}^{(r)}\right)\exp\left(-\boldsymbol{x}_i^T\boldsymbol{\alpha}^{(r)}\right)}{d\sum_{i=1}^{n}x_{ij}^2\exp\left(-\boldsymbol{x}_i^T\boldsymbol{\alpha}^{(r)}\right)}.$$

Next, we observe that (5) is linearly separable in $\alpha_j$ for $j=1,...,d$. Further, each $\alpha_j$ occurs within a linear composition inside of a negative exponential function, which implies that (5) is concave in $\boldsymbol{\alpha}$ (cf. [18, Exam. 3.13]). Unfortunately there is no closed-form solution for the first-order condition equation $\nabla Q_{\boldsymbol{\alpha}}\left(\boldsymbol{\alpha};\boldsymbol{\psi}^{(r)}\right) = \mathbf{0}$. However, we can compute

$$\boldsymbol{\alpha}^* = \arg\max_{\boldsymbol{\alpha}} Q_{\boldsymbol{\alpha}}\left(\boldsymbol{\alpha};\boldsymbol{\psi}^{(r)}\right)$$

by considering the partial derivative equations $(\partial/\partial\alpha_j) Q_{\boldsymbol{\alpha}}\left(\boldsymbol{\alpha};\boldsymbol{\psi}^{(r)}\right) = 0$ for each $j = 1, ..., d$ instead, where $\boldsymbol{\alpha}^{*T} = (\alpha_1^*, ..., \alpha_d^*)$. The solution to $(\partial/\partial\alpha_j) Q_{\boldsymbol{\alpha}}\left(\boldsymbol{\alpha};\boldsymbol{\psi}^{(r)}\right) = 0$ can be obtained via a Newton algorithm using the first and second partial derivatives

$$\frac{\partial Q_{\boldsymbol{\alpha}}}{\partial \alpha_j} = -\frac{1}{2}\sum_{i=1}^{n} x_{ij} + \frac{1}{2}\sum_{i=1}^{n} \frac{x_{ij}\left(y_i - \boldsymbol{x}_i^T\boldsymbol{\beta}^{(r)}\right)^2}{\exp\left(dx_{ij}\left(\alpha_j - \alpha_j^{(r)}\right) + \boldsymbol{x}_i^T\boldsymbol{\alpha}^{(r)}\right)}$$

and

$$\frac{\partial^2 Q_{\boldsymbol{\alpha}}}{\partial \alpha_j^2} = -\frac{1}{2}\sum_{i=1}^{n} \frac{x_{ij}^2\left(y_i - \boldsymbol{x}_i^T\boldsymbol{\beta}^{(r)}\right)^2}{\exp\left(dx_{ij}\left(\alpha_j - \alpha_j^{(r)}\right) + \boldsymbol{x}_i^T\boldsymbol{\alpha}^{(r)}\right)}.$$

Alternatively, a bisection algorithm can be used to obtain the root of each partial derivative equation; see for example [19, Sec. 9.1.1].

Using Propositions 1 and 2, the MM algorithm for computing (3) can be defined via the update scheme

$$\boldsymbol{\psi}^{(r+1)T} = \begin{cases} \left(\boldsymbol{\alpha}^*,\boldsymbol{\beta}^{(r)}\right) & \text{if } r \text{ is odd,} \\ \left(\boldsymbol{\alpha}^{(r)},\boldsymbol{\beta}^*\right) & \text{if } r \text{ is even,} \end{cases} \quad (7)$$

where $\boldsymbol{\alpha}^*$ and $\boldsymbol{\beta}^*$ are obtained via the descriptions above.

## 2.1 Convergence Analysis

Starting from some initial value $\boldsymbol{\psi}^{(0)}$, update scheme (7) is repeated until some numerical convergence criterion is reached; for example, the algorithm can be terminated once $\ell\left(\boldsymbol{\alpha}^{(r+1)}, \boldsymbol{\beta}^{(r+1)}\right) - \ell\left(\boldsymbol{\alpha}^{(r)}, \boldsymbol{\beta}^{(r)}\right) < \epsilon$ for some small $\epsilon > 0$. Upon termination, the final iterate of the algorithm is declared the ML estimator $\hat{\boldsymbol{\psi}}$. See [20, Sec. 11.5] regarding the relative merits of various convergence criteria.

Let $\boldsymbol{\psi}^{(\infty)} = \lim_{r \to \infty} \boldsymbol{\psi}^{(r)}$ (or alternatively, $\hat{\boldsymbol{\psi}} \to \boldsymbol{\psi}^{(\infty)}$ as $\epsilon < 0$) be a limit point of the blockwise MM algorithm. We have, from Propositions 1 and 2, that (5) and (6) are $\boldsymbol{\alpha}$- and $\boldsymbol{\beta}$-blockwise minorizers of (4), respectively. Further, both (5) and (6) are strictly concave and smooth in the respective parameter components. Thus, the MM algorithm satisfies the assumptions of [16, Thm. 2], which yields the following result.

**Proposition 3.** *Let $\boldsymbol{\psi}^{(r)}$ be a sequence of blockwise MM algorithm iterates (as defined by (7)) with limit $\boldsymbol{\psi}^{(\infty)}$, for some initial value $\boldsymbol{\psi}^{(0)}$. The following statements are true.*

**(a)** *The sequence of log-likelihood values $\ell\left(\boldsymbol{\alpha}^{(r)}, \boldsymbol{\beta}^{(r)}\right)$ is monotonically increasing in $r$.*

**(b)** *The limit point $\boldsymbol{\psi}^{(\infty)}$ is a stationary point of the log-likelihood function $\ell(\boldsymbol{\alpha}, \boldsymbol{\beta})$.*

A further result can be obtained by noting that (4) is biconcave in $\boldsymbol{\alpha}$ and $\boldsymbol{\beta}$. That is, for fixed $\boldsymbol{\alpha}$, (4) is concave since it consists of a concave quadratic function in $\boldsymbol{\beta}$. Similarly, for fixed $\boldsymbol{\beta}$, (4) is the sum of a linear function and the negative exponential composition of a linear function, which implies concavity in $\boldsymbol{\alpha}$ (cf. [18, Exam. 3.13]). Since (4) is also differentiable in $\boldsymbol{\psi}$, biconcavity implies the following result via [21, Thm. 4.2].

**Proposition 4.** *Every limit point $\boldsymbol{\psi}^{(\infty)}$ of the blockwise MM algorithm [as defined by (7)] is a coordinate-wise maximizer of $\ell(\boldsymbol{\alpha}, \boldsymbol{\beta})$ in both $\boldsymbol{\alpha}$- and $\boldsymbol{\beta}$-blocks (with the other block fixed).*

The monotonicity result from Proposition 3 guarantees that the blockwise MM algorithm is stable and will not take a step that decreases the objective log-likelihood value. We note that although Proposition 4 guarantees that the limit point is a coordinate-wise maximum of the log-likelihood, there is no theoretical guarantee that the limit point is a maximum and not a saddlepoint of (4).

## 3 A Parallel and Distributed Implementation

Suppose that we have a master processing element (PE) M and up to $d$ slave PEs $S_1$–$S_d$. Store $y_1, ..., y_n$ in each slave PE, and partition store the vector $x_{1j}, ..., x_{nj}$ on $S_j$ for $j = 1, ..., d$. Store an instance of the parameter vector $\boldsymbol{\psi}^{(0)}$ on the master PE and each of the slave PEs.

To initialize the algorithm, have each $S_j$ send M the quantities $x_{ij}\alpha_j^{(0)}$ and $x_{ij}\beta_j^{(0)}$ for each $i$. The master PE M then computes $\boldsymbol{x}_i^T\boldsymbol{\alpha}^{(0)}$ and $\boldsymbol{x}_i^T\boldsymbol{\beta}^{(0)}$ for each $i$, and sends the quantities to each of the slaves $S_j$.

At each odd iteration $r+1$, M sends $\alpha_j^{(r)}$, $\boldsymbol{x}_i^T\boldsymbol{\alpha}^{(r)}$, and $\boldsymbol{x}_i^T\boldsymbol{\beta}^{(r)}$ for each $i$, to each of the respective slave PEs $S_j$. Each $S_j$ then computes $\alpha_j^*$ and sends $\alpha_j^*$ and $x_{ij}\alpha_j^*$ for each $i$ to M. The master PE M then combines the quantities $\alpha_j^*$ and $x_{ij}\alpha_j^*$ to produce $\boldsymbol{\alpha}^{(r+1)}$ and $\boldsymbol{x}_i^T\boldsymbol{\alpha}^{(r+1)}$ for each $i$, respectively.

At each even iteration $r+1$, M sends $\beta_j^{(r)}$, $\boldsymbol{x}_i^T\boldsymbol{\alpha}^{(r)}$, and $\boldsymbol{x}_i^T\boldsymbol{\beta}^{(r)}$ for each $i$, to each of the respective slave PEs $S_j$. Each $S_j$ then computes $\beta_j^*$ and sends $\beta_j^*$ and $x_{ij}\beta_j^*$ for each $i$ to M. M then combines the quantities $\beta_j^*$ and $x_{ij}\beta_j^*$ to produce $\boldsymbol{\beta}^{(r+1)}$ and $\boldsymbol{x}_i^T\boldsymbol{\beta}^{(r+1)}$ for each $i$, respectively.

After initialization, the implementation requires the storage of only $2d + 2n$ real-valued quantities on M, and the storage of only $2 + 4n$ quantities on $S_j$ for each $j$, at any iteration $r > 0$. Furthermore, at each iteration $1 + 2n$ quantities are sent from M to each of the slaves $S_j$ and each slave sends $1 + n$ quantities back to M.

The algorithm that is described requires no matrix computations and allows for the data to be distributed between up to $d$ slave PEs. The role of each of the $d$ PEs can be partitioned over a smaller number of PEs if less than $d$ PEs are available. In [12], it is noted that such distributed algorithm are best implemented in parallel via graphics processing units; see also [22].

## 4 Time Complexity

Let $N_{n,\epsilon}^{\text{Newton}}$ be the average number of iterations required for the Newton algorithm for the computation of (2) [7, Sec. 11.7] to converge for $n$ observations and some criterion threshold $\epsilon$. Starting from some initial value $\boldsymbol{\psi}^{(0)}$, the $(r+1)$th iteration of the Newton algorithm, $\boldsymbol{\psi}^{(r+1)}$, requires the computation of the two steps

$$\boldsymbol{\beta}^{(r+1)} = \left[\sum_{i=1}^{n} \frac{\boldsymbol{x}_i \boldsymbol{x}_i^T}{\exp\left(\boldsymbol{x}_i^T \boldsymbol{\alpha}^{(r)}\right)}\right]^{-1} \sum_{i=1}^{n} \frac{\boldsymbol{x}_i y_i}{\exp\left(\boldsymbol{x}_i^T \boldsymbol{\alpha}^{(r)}\right)}$$

and

$$\boldsymbol{\alpha}^{(r+1)} = \boldsymbol{\alpha}^{(r)} - \left[\sum_{i=1}^{n} \boldsymbol{x}_i \boldsymbol{x}_i^T\right]^{-1} \sum_{i=1}^{n} \boldsymbol{x}_i$$
$$+ \left[\sum_{i=1}^{n} \boldsymbol{x}_i \boldsymbol{x}_i^T\right]^{-1} \frac{\sum_{i=1}^{n} \boldsymbol{x}_i \left(y_i - \boldsymbol{x}_i^T \boldsymbol{\beta}^{(r+1)}\right)^2}{\exp\left(\boldsymbol{x}_i^T \boldsymbol{\alpha}^{(r)}\right)}.$$

The Newton algorithm is derived by following the Fisher scoring formulation (cf. [20, Sec. 10.6]). We observe that both steps are dominated by the sums of $n$ outer produces of $d$ dimensional vectors, and $d \times d$ matrix inversions. The sum of products has order $O\left(nd^2\right)$ and the inversion has order $O\left(d^3\right)$; see [19, Secs. 2.1–2.3]. The overall order is thus $O\left(N_{n,\epsilon}^{\text{Newton}} \left[nd^2 + d^3\right]\right)$. Finally, like the blockwise MM algorithm, there is no theoretical guarantee that the Newton algorithm converges to a maximum of (4).

Let $N_{n,\epsilon}^{\text{MM}}$ be the average number of cycles (an odd and an even step) required for the blockwise MM algorithm to converge, for $n$ observations and some critical threshold $\epsilon$. From (7), we observe that when $r$ is even, the computation of $\boldsymbol{\beta}^*$ requires the computation of $\boldsymbol{x}_i^T\boldsymbol{\beta}^{(r)}$ and $\boldsymbol{x}_i^T\boldsymbol{\alpha}^{(r)}$ once, for each $i = 1, ..., n$, which requires $O(nd)$ operations. Given $\boldsymbol{x}_i^T\boldsymbol{\beta}^{(r)}$ and $\boldsymbol{x}_i^T\boldsymbol{\alpha}^{(r)}$, the computation of each $\beta_j^*$ requires $O(n)$ operations, for each $j = 1, ..., d$, thus the overall complexity is $O(nd)$ when $r$ is even. In each odd step, either a Newton algorithm or bisection algorithm is required to evaluate the roots of each of the $d$ equations $(\partial/\partial\alpha_j) Q_{\boldsymbol{\alpha}}(\boldsymbol{\alpha}; \boldsymbol{\psi}^{(r)}) = 0$. To solve these equations, $\boldsymbol{x}_i^T\boldsymbol{\beta}^{(r)}$ and $\boldsymbol{x}_i^T\boldsymbol{\alpha}^{(r)}$ are required to be computed once, for each $i$. Let $N_{n,\epsilon}^{\text{Root}}$ be the number of iterations required by the root-finding algorithm. Given $\boldsymbol{x}_i^T\boldsymbol{\beta}^{(r)}$ and $\boldsymbol{x}_i^T\boldsymbol{\alpha}^{(r)}$, the dominant term in each root-finding algorithm iteration is dominated by $n$ times a constant number of operations, for each of the $d$ components of $\boldsymbol{\alpha}^*$. Therefore, each odd step has complexity order $O\left(nd + N_{n,\epsilon}^{\text{Root}} nd\right)$. The overall order is thus $O\left(N_{n,\epsilon}^{\text{MM}} \left[nd + N_{n,\epsilon}^{\text{Root}} nd\right]\right)$.

## 5 Simulation Studies

We now report on a set of simulation studies. In our simulation studies, we generate a sample of $n = 100, 1000, 10000$ observations from the model $\phi\left(y; \boldsymbol{\beta}^T\boldsymbol{x}, \exp\left[\boldsymbol{\alpha}^T\boldsymbol{x}\right]\right)$, where $d \in \{5, 10, 20, 50\}$ in all cases of $n$. Here $10\alpha_j$, $\beta_j$, and $x_{ij}$ are each randomly generated from a standard normal distribution. Using each sample, we compute $\hat{\boldsymbol{\psi}}$ via the MM algorithm and the Newton algorithm of [7, Sec. 11.7]. The process is repeated 100 time; the computation time and convergence status of the algorithm is recorded from each repetition. The average and standard deviation of computation times, are reported in Table I. Also reported in Table I is the theoretical computation time of the MM algorithm under parallelization, which is the computation time divided by $d$, where $d$ is the maximum possible number of slave PEs that can be used. The theoretical computation time under parallelization assumes negligible communication times between PEs.

Table 1: Average computation times (over 100 replications) are presented in in boldface. Standard deviations are presented in italics. *No replication of the Newton algorithm converged.

|  |  |  | $d=$ | | | |
|---|---|---|---|---|---|---|
|  |  |  | 5 | 10 | 20 | 50 |
| $n=$ | 100 | Newton | **0.0007** | **0.0025** | **0.0804** | —* |
|  |  |  | *0.0005* | *0.0009* | *0.3682* | —* |
|  |  | MM (Serial) | **0.0028** | **0.0227** | **0.6155** | **4.7669** |
|  |  |  | *0.0009* | *0.0068* | *0.3528* | *1.4667* |
|  |  | MM (Parallel) | **0.0003** | **0.0021** | **0.0546** | **0.4246** |
|  |  |  | *0.0002* | *0.0018* | *0.0487* | *0.0003* |
|  | 1000 | Newton | **0.0280** | **0.0550** | **0.1222** | **0.4846** |
|  |  |  | *0.0022* | *0.0061* | *0.0090* | *0.0326* |
|  |  | MM (Serial) | **0.0153** | **0.0908** | **0.5052** | **7.0412** |
|  |  |  | *0.0023* | *0.0099* | *0.0447* | *0.4897* |
|  |  | MM (Parallel) | **0.0014** | **0.0084** | **0.0468** | **0.6527** |
|  |  |  | *0.0011* | *0.0064* | *0.0348* | *0.4812* |
|  | 10000 | Newton | **3.9243** | **5.5251** | **10.7438** | **30.0419** |
|  |  |  | *0.3181* | *0.1901* | *1.0519* | *3.3048* |
|  |  | MM (Serial) | **0.1665** | **0.8654** | **4.7978** | **64.5258** |
|  |  |  | *0.0172* | *0.0574* | *0.2067* | *4.9660* |
|  |  | MM (Parallel) | **0.0154** | **0.0806** | **0.4460** | **5.9394** |
|  |  |  | *0.0116* | *0.0607* | *0.3356* | *4.3808* |

The algorithms were applied via implementations in the $R$ programming environment ([23]; version 3.2.2) on an Intel Core i7 CPU running at 2.40 GHz with 16 GB internal RAM, and the timing was conducted using the *proc.time* function. Furthermore, all mathematical functions are programmed in $C$ and integrated via *Rcpp* and *Rcpparmadillo* [24]. The MM algorithm is thresholded using the convergence criterion described in Section II.A and the constant $\epsilon = 10^{-3}$. An absolute convergence criterion is used for all Newton algorithms with a threshold $\epsilon = 10^{-3}$; see [20, Sec. 11.5].

## 5.1 Results

There are a number of notable features from Table I. Firstly, we note that the Newton algorithm could not be implemented on the relatively small data case of $d = 50$ and $n = 100$. Upon inspection, the Newton algorithm suffers from problems of rank deficiencies in matrix inversions, causing the log-likelihood values to diverge.

Secondly, for all $d$ in the $n = 100$ case (where comparable), the parallel MM algorithm is faster than the Newton algorithm, which is faster than the serial MM algorithm. This trend changes in the $n = 1000$ case where the serial MM algorithm is faster than the Newton algorithm when $d = 5$. In the $n = 10000$ case, we observe that the serial MM algorithm is faster than the Newton algorithm when $d = 5, 10, 20$. We conclude that the serial implementation of the MM algorithm is faster than the Newton algorithm when $d$ is small and $n$ is large. Furthermore, the MM algorithm is more stable and can be applied where the Newton algorithm may fail.

In Figure 1, we plot the average log-ratios of computation times between the serial MM and Newton algorithms, and the theoretical parallel MM and Newton algorithms. We observe that the serial implementation of the MM algorithm can have computation time as fast as $< 1/23$ times that of the Newton algorithm ($d = 5$, $n = 10000$) or as slow as $> 15$ times that of the Newton algorithm ($d = 50$, $n = 1000$), on average. The parallel implementation can have computation time as fast as $< 1/254$ times that of the Newton algorithm ($d = 5$, $n = 10000$) or as slow as $> 1.4$ times that of the Newton algorithm ($d = 50$, $n = 1000$) on average.

We further observe that the computation time ratios of the MM algorithm to the Newton Algorithm, in both serial and parallel, are decreasing in $n$ and increasing in $d$. Thus, we suggest that the MM algorithm is preferable to the

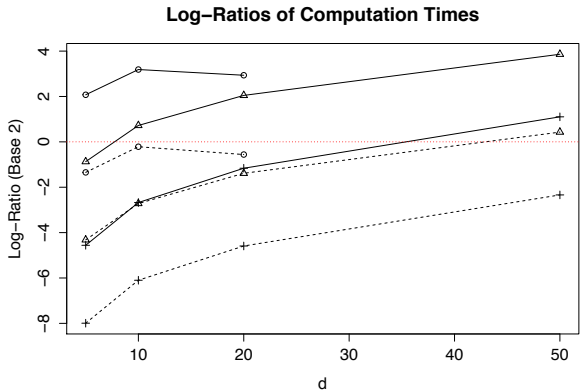

Figure 1: Average log-ratios (base 2) are plotted. Ratios between the serial MM and Newton algorithms are presented as solid lines. Ratios between the parallel MM and Newton algorithms are presented as dashed lines. Circles, triangles, and pluses indicate $n = 100, 1000, 10000$, respectively.

Newton algorithm in cases where $n$ is large and $d$ is relatively small.

## 6  Conclusions

In this letter, we introduce an MM algorithm for the computation of (2) that requires no matrix operations. The algorithm is shown to be globally convergent and to generate monotonic sequences of log-likelihood values. Furthermore, a distributed and parallel implementation of the MM algorithm is described and it is shown that the MM algorithm has a different order of computational complexity to the Newton algorithm.

Via simulation studies, the serial implementation is demonstrated to have computation time as fast as $< 1/23$ and the parallel implementation is hypothesized to have computation time $< 1/254$ times that of the Newton algorithm, when $n$ is large and $d$ is relatively small. We thus recommend the use of the MM algorithm in such scenarios.